\documentclass[english,aip,apl,reprint]{revtex4-1}

\usepackage{graphicx}
\usepackage{bm}
\usepackage{natbib} 
\usepackage{color}

\begin{document}

\title{\textcolor{blue}{Characterization and Suppression of Low-frequency Noise in Si/SiGe Quantum Point Contacts and Quantum Dots}}

\author{K.~\surname{Takeda}}
\email{takeda@meso.t.u-tokyo.ac.jp}
\affiliation{Department of Applied Physics, The University of Tokyo, Hongo, Bunkyo-ku, Tokyo, 113-8656, Japan}

\author{T.~\surname{Obata}}
\affiliation{Department of Applied Physics, The University of Tokyo, Hongo, Bunkyo-ku, Tokyo, 113-8656, Japan}

\author{Y.~\surname{Fukuoka}}
\affiliation{Quantum Nanoelectronics Research Center, Tokyo Institute of Technology, O-okayama, Meguro-ku, Tokyo 152-8552, Japan}

\author{W. M.~\surname{Akhtar}}
\affiliation{Department of Applied Physics, The University of Tokyo, Hongo, Bunkyo-ku, Tokyo, 113-8656, Japan}

\author{J.~\surname{Kamioka}}
\affiliation{Quantum Nanoelectronics Research Center, Tokyo Institute of Technology, O-okayama, Meguro-ku, Tokyo 152-8552, Japan}

\author{T.~\surname{Kodera}}
\affiliation{Quantum Nanoelectronics Research Center, Tokyo Institute of Technology, O-okayama, Meguro-ku, Tokyo 152-8552, Japan}
\affiliation{PRESTO, Japan Science and Technology Agency (JST), Honcho Kawaguchi, Saitama, Japan} 

\author{S.~\surname{Oda}}
\affiliation{Quantum Nanoelectronics Research Center, Tokyo Institute of Technology, O-okayama, Meguro-ku, Tokyo 152-8552, Japan}

\author{S.~\surname{Tarucha}}
\affiliation{Department of Applied Physics, The University of Tokyo, Hongo, Bunkyo-ku, Tokyo, 113-8656, Japan}
\affiliation{RIKEN, Center for Emergent Matter Science (CEMS), Wako, Saitama, 351-0198, Japan}

\date{\today}

\begin{abstract}
We report on the effects of a global top gate on low-frequency noise in Schottky gate-defined quantum point contacts (QPCs) and quantum dots (QDs) in a modulation-doped Si/SiGe heterostructure. 
For a relatively large top gate voltage, the QPC current shows frequent switching with $1/f^2$ Lorentzian type charge noise. 
As the top gate voltage is decreased, the QPC pinch-off voltage becomes less negative, and the $1/f^2$ noise becomes rapidly suppressed in a homogeneous background $1/f$ noise. 
We apply this top-gating technique to double QDs to stabilize the charge state for the electron number down to zero.
\end{abstract}


\maketitle

Coherent control of single electron spin in solids aiming towards the realization of fundamental devices for quantum computation has been performed extensively in GaAs\cite{Petta_Science2005, Koppens_Nature2006, Nowack_Science2007, MPL_NatPhys2008, Roland_PRL2011, Shulman_Science2012} and more recently in Si\cite{Maune_SiGe_ST_Nature2012, Pla_Nature2012} as well.
Si quantum dots (QDs) are one of the most promising candidates for implementing scalable spin quantum bit (qubit) systems because of the long coherence time due to weak spin-orbit and hyperfine interactions.
Recent experiments on Si QDs have shown the relaxation time $T_1$ and the dephasing time $T_2 ^*$ much longer than those observed for GaAs QDs.\cite{Simmons_PRL2011, Prance_PRL2012, Maune_SiGe_ST_Nature2012}
However, at present the realization of stable QDs in a two-dimensional electron gas (2DEG) at a modulation-doped Si/SiGe heterostructure is still challenging due to the problem of charge noise which cause sudden changes to the QD states.\cite{Wild_NJP2010, Payette_LSRL_APL2012} 
To realize stable qubit operation it is necessary to characterize and reduce the charge noise in quantum point contacts (QPCs) which form tunnel junctions with the QD.

Charge noise in gate-defined QPCs has been studied in detail for modulation-doped GaAs/AlGaAs heterostructures using techniques of asymmetrically biasing gates\cite{Sakamoto_APL1995}, 
bias-cooling\cite{Michel_chargenoise_PRB2005}, 
top gate biasing\cite{Christo_chargenoise_PRL2008} and combinations of them. 
In Ref. \cite{Sakamoto_APL1995} they observed charge noise caused by thermally activated trapping and detrapping by charge traps in local potential minima formed near the QPC channel when Schottky gates were occasionally biased to align the energy level of the charge trap and Fermi level of the QPC channel.
On the other hand, in Ref. \cite{Michel_chargenoise_PRB2005, Christo_chargenoise_PRL2008}, the charge noise is attributed to the electron tunneling between the surface and the 2DEG 
 via charge traps in between.
They observed strong charge noise reduction by application of bias-cooling or top gate technique which makes the operation voltage of the surface Schottky gate less negative. Then the tunnel rate of electrons from the Schottky gate to the QPC channel via charge traps or most probably ionized donor states (discussed later) is reduced to improve the charge stability.
For Si/SiGe Schottky gated devices, both mechanisms can also be considered as noise sources.

The top gate-biasing may be useful to reduce the charge noise in Si/SiGe devices, but not the bias cooling because $n$-type donors in SiGe do not have deep levels like DX centers \cite{Buks_DXcenter_1994} in $n$-doped AlGaAs. 
It has been demonstrated for Si/SiGe QPCs defined by gate electrodes placed on a Al$_2$O$_3$ insulated Si surface, i.e., metal-oxide-semiconductor (MOS), that negatively biasing a global top gate gradually reduces the charge noise in the conductance.\cite{Wild_APL2012}
The charge noise may be more significant in surface Schottky gate structures than in MOS gate structures because there is no direct leakage between the gate metal and the QPC channel in the latter, 
but no detailed studies on charge noise spectra in Si/SiGe QPCs have been reported to date. 

In this letter, we characterize low-frequency charge noise spectra in Schottky gated Si/SiGe QPCs with a global top gate voltage.
We observe a strong reduction in the $1/f^2$ component in the charge noise by negatively biasing the top gate. 
We then apply this technique to enable stable operation of double QDs and finally achieve stable charge state with just a few electrons.

\begin{figure}[t]
\centering
\includegraphics*[width = 1.00\columnwidth]{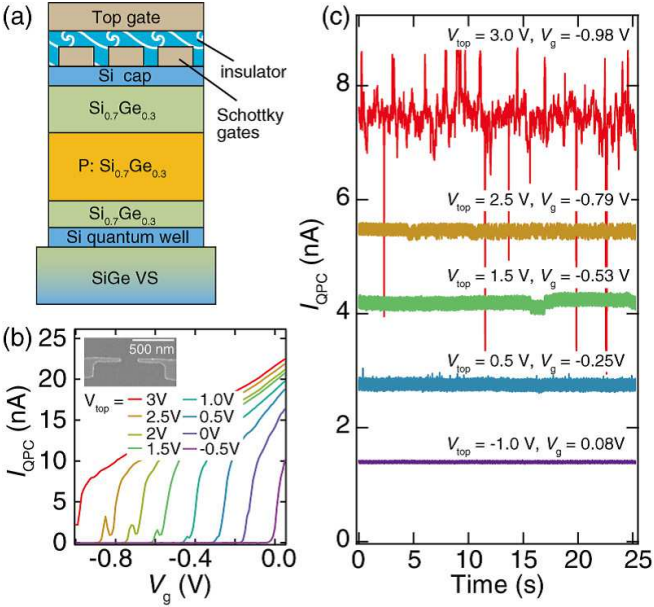}
\caption{\label{Fig1} (a) Layer sequence of the Si/SiGe heterostructure and gate electrodes used in this study. 
The Pd Schottky gate is deposited on the Si cap followed by deposition of the insulator and top gate metal.
 (b) QPC current $I_{\mathrm{QPC}}$ as a function of Schottky gate voltage $V_g$ for various top gate voltages $V_{\mathrm{top}}$ ($T=4.2$ K, $V_{\mathrm{SD}} = 500$ $\mathrm{\mu V}$). 
Pinch-off point of the QPC can be tuned by application of top gate voltage. Inset: scanning electron micrograph of the sample.
 (c) QPC current time traces for various top gate voltages offset for clarity. The sampling rate is 1 kHz.}
\end{figure}

The Si/SiGe heterostructure used in this study is grown by ultra-high vacuum chemical vapor deposition. 
The heterostructure configuration is shown in Fig. 1(a). First a 3 $\mathrm{\mu m}$ thick graded buffer is grown on a Si substrate
by linearly increasing the Ge content from 0 \% to 30 \%, and then a 1 $\mathrm{\mu m}$ thick Si$_{0.7}$Ge$_{0.3}$ buffer is grown. 
A 15 nm thick Si quantum well, a 10 nm thick undoped Si$_{0.7}$Ge$_{0.3}$ spacer, a 25 nm thick phosphorous doped Si$_{0.7}$Ge$_{0.3}$, a 20 nm thick undoped Si$_{0.7}$Ge$_{0.3}$ cap, and finally a 7.5 nm thick Si cap are successively grown on the buffer. 
The resulting 2DEG has an electron density of 2.3 $\times$ 10$^{11}$  cm$^{-2}$ and a mobility of 1.3 $\times$ 10$^{5}$ cm$^2$/Vs at $T=250$ mK.

The stack of gate electrodes is also shown in Fig. 1(a).
Ohmic contacts to the 2DEG are achieved by Au/Sb evaporation and subsequent annealing. 
QPCs are formed by Pd Schottky gates on the Si cap layer surface by a standard electron-beam lithography technique. 
The top gate which covers the whole active QPC area is placed on top of the surface gates with cross-linked PMMA as an insulator in between. 

Figure 1(b) shows the QPC current $I_{\mathrm{QPC}}$ v.s. surface gate voltage $V_{\mathrm{g}}$ measured for various top gate voltages $V_{\mathrm{top}}$ at $T=4.2$ K. 
The QPC pinch-off voltage becomes less negative as $V_{\mathrm{top}}$ is decreased. 
Figure 1(c) shows $I_{\mathrm{QPC}}$ time traces measured for various $V_{\mathrm{top}}$. 
The Schottky gate is biased at a condition where the slope of QPC current  ($\mathrm{d}I_{\mathrm{QPC}}/\mathrm{d}V_g$) is largest to maximize the charge sensitivity. 
It is clear that both magnitude and frequency of the current fluctuation become significantly small as the top gate is made less positive or negative. 
The minimum top gate voltage that we can apply in this sample is $V_{\mathrm{top}}=-1$ V, because the QPC conductance rapidly quenches for $V_{\mathrm{top}} < -1$ V.

Next we analyze power spectra of the charge noise using fast Fourier transform (FFT) of $I_{\mathrm{QPC}}$ time traces. 
Figure 2(a) shows FFT power spectra $S_I (f)$ of $I_{\mathrm{QPC}}$ in the low-frequency range of $f=0.01$ to 10 Hz measured for various $V_{\mathrm{top}}$.
The  $S_I (f)$ shows a $1/f^2$ behavior known as two-level Lorentzian type fluctuation \cite{noise_book} at relatively large $V_{\mathrm{top}}$. 
In this condition charge fluctuations due to just a few charge traps 
near the QPC channel are dominant in the noise characteristic. 
On the other hand, as the $V_{\mathrm{top}}$ is made less positive, the $1/f^2$ noise is rapidly suppressed and disappear in the $1/f$ type homogeneous background charge noise which comes from summation of many trapping sites. \cite{noise_book} 
To more quantitatively characterize the noise feature, we analyze the integrated spectral density over a finite frequency range to derive the mean value of the gate voltage fluctuation amplitude, $\Delta V_{\mathrm{EG}}$ of the noise as in ref \cite{Christo_chargenoise_PRL2008, Jung_APL2004}

\begin{equation}
\Delta V_{\mathrm{EG}} = \sqrt{2 \int _{0.01} ^{49} [S_I (f) - S_{I; BG} (f) ]df} \bigg/ \biggl( \frac{dI_{\mathrm{QPC}}}{dV_{\mathrm{G}}} \biggr) .
\end{equation}

\begin{figure}[t]
\centering
\includegraphics*[width=1.00\columnwidth]{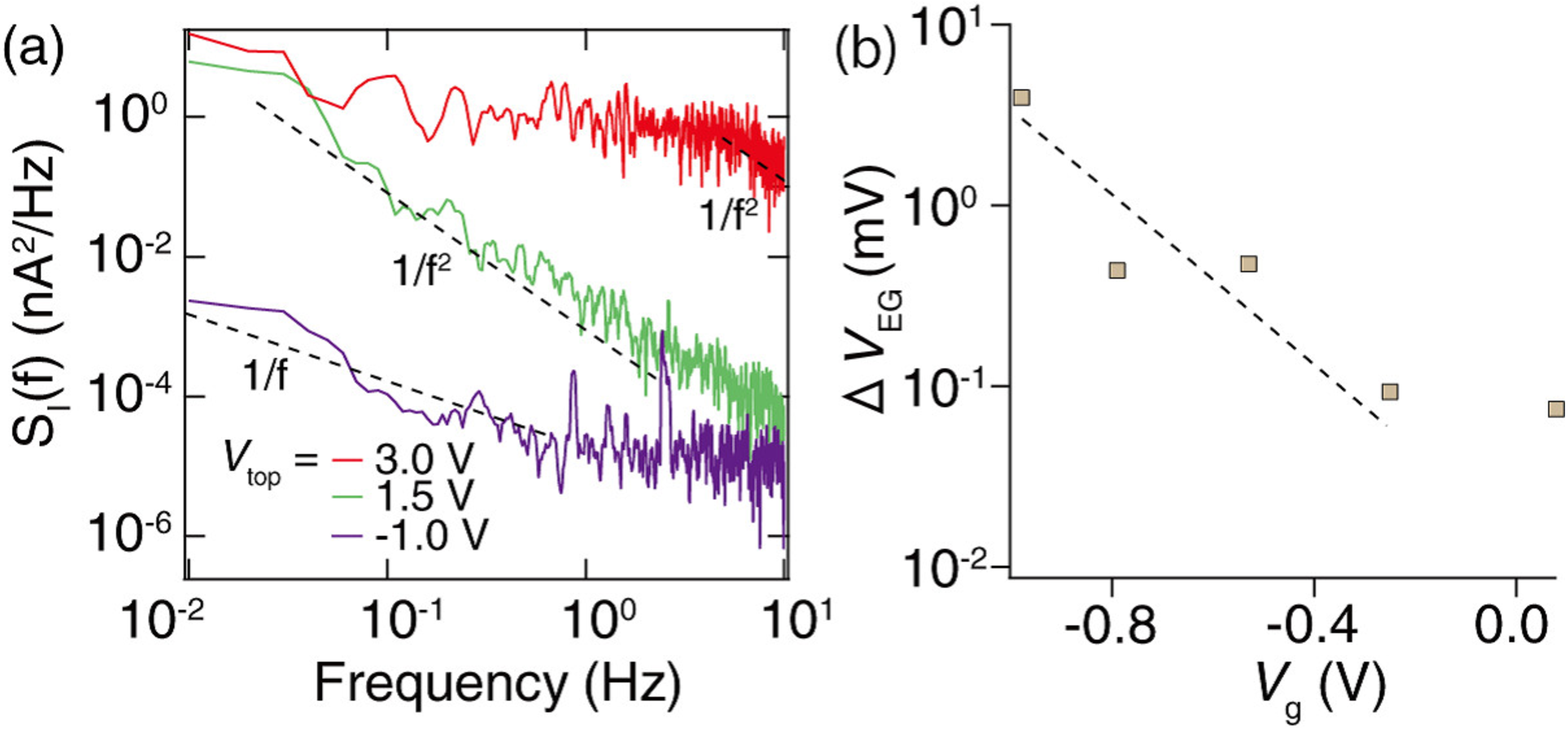}
\caption{\label{Fig2} (a) Power spectra $S_I (f)$ of the charge noise obtained by FFT.  Time traces used for FFT are measured with a 100 Hz sampling rate for 250 sec. 
(b) Mean value of the gate voltage fluctuation amplitude $\Delta V_{\mathrm{EG}}$ of the noise obtained by normalizing the experimental data by the sensitivity of the QPC at each top gate voltage and Schottky gate voltage.}
\end{figure}

Here $S_{I; BG} (f)$ is the background noise measured at $V_g = V_{\mathrm{top}} = V_{\mathrm{SD}} = 0$ V.
We derive it for a frequency range of 0.01 to 49 Hz to avoid a relatively large 50 Hz noise which comes from measurement instruments.
The obtained $\Delta V_{\mathrm{EG}}$ shown in Fig. 2(b) changes exponentially as a function of $V_g$.
This is similar to that in a previous report on GaAs QPCs, but the residual $\Delta V_{\mathrm{EG}}$ seems nearly one order larger than that of GaAs QPCs.\cite{Christo_chargenoise_PRL2008}
We believe further reduction of the charge noise level can be achieved by optimization of the heterostructure and gating technique.

\begin{figure*}[!t]
\centering
\includegraphics*[width=2.00\columnwidth]{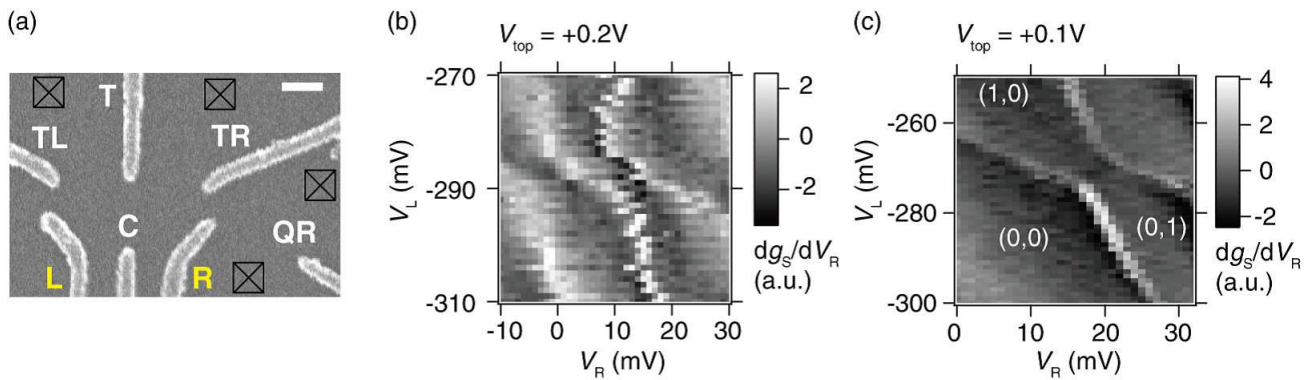}
\caption{\label{Fig3} (a) Scanning electron micrograph of the double quantum dot device similar to the one used in the measurement.  The scale bar corresponds to 100 nm and black boxes represent Ohmic contacts to the 2DEG.
(b), (c) Charge stability diagram \cite{SpinsInFewElectron} of few-electron double quantum dot. Other gates are biased at $V_{\mathrm{T}}=+0.1$ V, $V_{\mathrm{TR}}=+0.25$ V, 
$V_{\mathrm{TL}}=-0.3$ V, $V_{\mathrm{C}}=+0.35$ V, and $V_{\mathrm{QR}}=+0.165$ V for both cases. ($T=250$ mK)
: Noisy stability diagram measured at $V_{\mathrm{top}}=+0.2$ V (b), and 
quiet charge stability diagram measured at $V_{\mathrm{top}}=+0.1$ V (c).}
\end{figure*}

The charge noise reduction observed in our Si/SiGe system can be explained by considering surface - 2DEG tunneling which has already been confirmed in GaAs devices.\cite{Michel_chargenoise_PRB2005, Christo_chargenoise_PRL2008} 
In usual depletion type devices, application of less positive or more negative voltage to the top gate makes less negative the surface Schottky gate voltage to pinch off the QPC channel, and therefore the gate leakage current is reduced due to reduced electron tunneling between the surface and the channel. 
We exclude trapping and detrapping of charge traps near the QPC channel  from the reason for the charge noise 
because the observed charge noise decreases monotonically, but not resonantly 
when the top gate voltage is reduced.
It may be reasonable to assign the charge traps to modulation doped ionized impurities because recent report on an undoped Si/SiGe heterostructure has shown a significant improvement in the charge stability.\cite{Borselli_APL2011}

Finally we prepare a double QD in the same Si/SiGe heterostructure defined by the same gating technique and measure the stability diagram using the top gate voltage as one parameter. 
Figure 3(a) shows the scanning electron micrograph of the double QD. 
The charge state of the double QD is monitored using a proximal QPC as a charge sensor. 
The measured stability diagram in the plane of two side gate voltages $V_{\mathrm{R}}$ and $V_{\mathrm{L}}$ is shown in Fig. 3 measured at $V_{\mathrm{top}}=0.2$ V (b)
 and $V_{\mathrm{top}}=0.1$ V (c), respectively. 
The gray scale indicates the numerical derivative $\mathrm{d}g_{\mathrm{S}}/\mathrm{d}V_{\mathrm{R}}$ of the charge sensor conductance $g_{\mathrm{S}}$ with $V_{\mathrm{R}}$. 
The surface gate geometry is designed to have a relatively small gap in each QPC so that only a small positive voltage is applied to the top gate to form a double QD. 
The double QD is depleted to the (0, 0) charge configuration by adjusting $V_{\mathrm{R}}$ and $V_{\mathrm{L}}$. 
The (0, 0) state is confirmed because any extra charge transition line is observed for more negative gate voltages. 
In Fig. 3(b) charge state transition lines between the dot and 2DEG leads and between two dots are unclear due to electrostatic potential fluctuations by charge noise. 
In contrast, in Fig. 3(c) all charge state transition lines are better defined, while the operation voltages of the Schottky gates are not so different between Fig 3 (b) and Fig 3 (c). 
Note that the surface to 2DEG tunneling via charge traps can be reduced even without significant changes in the surface Schottky gate voltage because the energy level of the charge traps relative to the Fermi level of the gate metal can be modified just by application of the top gate voltage.

In summary, we have fabricated surface Schottky gate defined QPCs and double QDs in Si/SiGe heterostructures with a global top gate to study the top gate effect on the charge noise. 
We measured the real time current noise and noise power spectra for the QPCs and observed that $1/f^2$ switching noise is rapidly suppressed and finally disappear in the $1/f$ noise background as the top gate voltage is made less positive or more negative. 
This result is similar to that reported for clean GaAs QPCs\cite{Christo_chargenoise_PRL2008}. 
For the double QD, we also observed a similar effect of the top gate to stabilize the charge state with reduced charged noise. 
We could finally achieve a few-electron double QD with no notable switching noise. 
This top gate technique will be useful to stabilize the gate performance in Si/SiGe QDs 
and eliminate dephasing of qubits due to charge noise via exchange interaction. \cite{Culcer_APL2009}\\

We gratefully thank Juergen Sailer, Andreas Wild, Dominique Bougeard, and Gerhard Abstreiter for helpful discussions.
This work was financially supported by 
GCOE for Physical Sciences Frontier, MEXT, Japan, 
Project for Developing Innovation Systems of the Ministry of Education, Culture, Sports, Science and Technology, MEXT, Japan, 
Grant-in-Aid for Scientific Research on Innovative Areas (21102003), MEXT, Japan, 
and Funding for World-Leading Innovative R\&D on Science and Technology (FIRST) Program, Japan.

\end{document}